\documentclass[draft]{agujournal2019}
\usepackage{url} 
\usepackage{lineno}
\usepackage{amsmath}
\usepackage[inline]{trackchanges} 
\usepackage{soul}
\draftfalse

\journalname{Journal of Geophysical Research: Atmospheres}

\begin{document}

%
%


\title{A Satellite Remote Sensing and Doppler LiDAR-based Framework for Evaluating Mesoscale Flows Driven by Surface Heterogeneity}

%
%




\authors{T. Waterman\affil{1,2}, P. Germ\affil{2}, M. Calaf\affil{1}, E. Pardyjak\affil{1}, N. Chaney\affil{2}}

\affiliation{1}{Mechanical Engineering, University of Utah, Salt Lake City, UT, USA}
\affiliation{2}{Civil and Environmental Engineering, Duke University Pratt School of Engineering, Durham, NC, USA}




\correspondingauthor{Tyler Waterman}{tyswater@gmail.com}



\begin{keypoints}
\item Introduces a novel method combining satellite-derived land surface temperature and Doppler LiDAR data to detect mesoscale atmospheric flows.
\item Demonstrates the effectiveness of Dispersive Kinetic Energy (DKE) as an indicator of heterogeneity-driven flows in the boundary layer.
\item Shows DKE is correlated with satellite-based heterogeneity parameters, especially when controlling for the influence of large scale flows.
\end{keypoints}

%
%

%
%


\begin{abstract}
Surface heterogeneity, particularly complex patterns of surface heating, significantly influences mesoscale atmospheric flows, yet observational constraints and modeling limitations have hindered comprehensive understanding and model parameterization. This study introduces a framework combining satellite remote sensing and Doppler LiDAR to observationally evaluate heterogeneity-driven mesoscale flows in the atmospheric boundary layer. We quantify surface heterogeneity using metrics derived from GOES land surface temperature fields, and assess atmospheric impact through the Dispersive Kinetic Energy (DKE) calculated from a network of Doppler LiDAR profiles at the Southern Great Plains (SGP) Atmospheric Radiation Measurement (ARM) site. 

Results demonstrate that DKE and its ratio to the Mean Kinetic Energy (MKE) serve as effective indicators of heterogeneity driven flows, including breezes and circulations. The DKE and DKE ratio are correlated with metrics for surface heterogeneity, including the spatial correlation lengthscale, the spatial standard deviation, and the orientation of the surface heating gradient relative to the wind. The correlation becomes stronger when other flows that would affect DKE, including deep convection, low level jets, and storm fronts, are accounted for. Large Eddy Simulations contextualize the findings and validate the metric’s behavior, showing general agreement with expectations from prior literature. Simulations also illustrate the sensitivity to configuration of LiDAR networks using virtual LiDAR sites, indicating that even smaller networks can be used effectively. This approach offers a scalable, observationally grounded method to explore heterogeneity-driven flows, advancing understanding of land-atmosphere interactions as well as efforts to parameterize these dynamics in climate and weather prediction models.
\end{abstract}

\section*{Plain Language Summary}
Land surfaces are not uniform with some areas heating up faster than others, creating temperature differences that can drive changes in the atmosphere, including causing breezes to occur between warm and cool areas. These heterogeneity-driven flows affect weather and climate but are hard to observe and often missed by large-scale models. This study presents a new way to look at these flows using satellite and ground-based instruments. We use temperature maps from weather satellites to measure how uneven the surface heating is, and Doppler LiDAR sensors to track wind patterns above the ground. By analyzing how wind varies across space and time, we calculate a metric called Dispersive Kinetic Energy (DKE), which helps identify difficult to capture circulations. We tested this methodology at a research site in Oklahoma and compared the results with simulations. The findings show that our approach may detect these flows, opening up new opportunities to study how the land and atmosphere interact when the land is non uniform. Further results could be used to help improve weather and climate models in the future.

%
%

%


%
%
%
%

\section{Introduction} \label{sec:intro}
A wealth of literature from the past decades shows that surface spatial heterogeneity can have a wide variety of impacts on surface fluxes, the atmospheric boundary layer (ABL), and the broader atmosphere. Of particular interest in large scale models used to study weather and climate is small, mesoscale flows driven by differences in surface heating. In this context, with surface heating patches on the scale of kilometers, two processes in particular are well established to impact the ABL and above: the development of internal boundary layers, and heterogeneity driven circulations or roll structures, often called secondary circulations \cite{bou-zeid_persistent_2020}. These flows have been shown to increase regional fluxes, cloud development and  turbulent kinetic energy (TKE) \cite{avissar_three-dimensional_1996,weaver_coupling_2004,simon_heterogeneous_2024,simon_semicoupling_2021,zheng_impacts_2021,zhang_largeeddy_2023,paleri_impact_2025} while causing errors in point, tower-based measurement of surface fluxes \cite{prabha_characteristics_2007,patton_influence_2005,paleri_impact_2025}. As the impacts of surface heterogeneity on the atmosphere have become increasingly clear, more attention has been placed in developing parameterizations of heterogeneity driven flows for Earth System Models (ESMs) used to study climate processes and Numerical Weather Prediction (NWP) where the atmospheric grid resolution is often insufficient to resolve these flows.

Heterogeneity driven flows, especially secondary circulations, have primarily been studied using Large Eddy Simulation (LES) experiments where the flows can be resolved. LES experiments have examined the problem from small to large scales of heterogeneity, with idealized and (less often) realistic surface patterns, and under a variety of background conditions. In these studies, differential surface heating causes pressure differences in the near-surface atmosphere, flow then converges over lower density regions causing updrafts which then impose an inverse temperature gradient aloft to cause a complete circulation \cite{rochetin_morphology_2017}. This can alter not only regional advection, but also vertical transport as regional, dispersive heat fluxes driven by heterogeneity have been shown to reach up to 10\% of the turbulent heat flux, and 25\% of the moisture flux, even with relatively weak heterogeneity \cite{paleri_impact_2025}. Together, within the LES framework, these studies have identified a number of key parameters that control heterogeneity driven flows or modulate their broader impact on the atmosphere. Studies identify that both the magnitude of the surface heterogeneity (for example, difference in heating) and the spatial scale of the patches as important \cite{avissar_evaluation_1998,han_response_2019, lee_effect_2019,margairaz_surface_2020}. The length of the surface spatial pattern relative to the length of the boundary layer is also key \cite{patton_influence_2005,margairaz_surface_2020,van_heerwaarden_scaling_2014}. There is a minimum lengthscale necessary for notable large scale response that appears to be on the order of the boundary layer height, although there is some dissagreement on the exact relationship. Some studies estimate that spatial heterogeneity on the scale of 4-9 times the boundary layer height is optimal for secondary circulations \cite{patton_influence_2005}. The magnitude of the background wind is well established as a critical factor, with large wind speeds blending the surface and preventing atmospheric impact, although there is some disagreement on the maximum velocity that still allows for heterogeneity driven flows \cite{avissar_evaluation_1998,eder_secondary_2015, maronga_large-eddy_2013,rochetin_morphology_2017,weaver_coupling_2004}. The impact that the background wind has in suppressing heterogeneity driven flows is highly dependent on the orientation of the surface heterogeneity relative to the background wind; when they are perpendicular and lengthscales of heterogeneity are significant, horizontal breezes can form between warm and cool patches with even rather large ($>15 \ ms^{-1}$) geostrophic wind speeds \cite{weaver_coupling_2004,zhang_largeeddy_2023,prabha_characteristics_2007,rochetin_morphology_2017}. Other studies have suggested additional controlling factors, including atmospheric stability \cite{paleri_impact_2025}. 

Despite the wealth of information from LES, much remains unknown, uncertain, and challenging to ascertain about how surface heterogeneity impacts atmospheric flows. For real surfaces, defining key parameters of surface heterogeneity can be challenging as surface heterogeneity is highly multiscale. Even if the critical scale(s), orientation and spatial configuration of the surface could be well described simply, spatial patterns do not always persist in time \cite{torresrojas_geostatisticsbased_2024} which would have obvious impacts that need to be explored. The majority of LES studies have either idealized configurations (checkerboard, two-patch), persistent surface patterns (only change in magnitude, if at all, through time) or both which other studies of spatio-temporal patterns of LST across CONUS have shown to be relatively rare for a full diurnal cycle outside of permanent structural heterogeneity (land sea boundaries, hillslopes, urban boundaries)  \cite{torresrojas_geostatisticsbased_2024}. The orientation and magnitude of the background, geostrophic wind will also have a spatiotemporal variability that will affect how the surface patterns are translated to the atmosphere. Additionally, surface heterogeneity and its impact can be non-local. In LES studies which employ periodic boundary conditions, for example, patterns of heterogeneity can be unduly magnified as upstream heterogeneity is essentially assumed to be identical to local heterogeneity \cite{simon_heterogeneous_2024} whereas in real world conditions, upstream heterogeneity may be different with an alternative imprint on the local atmosphere for the domain of interest \cite{papangelis_saharan_2021}. Finally, while the framing of this work and many of these studies focuses on the impact the land has on the atmosphere, there is a dynamic, two-way coupling at play. The atmosphere can also impact the land via precipitation, cloud shading and non-linear blending from strong velocities that may vary in both direction and magnitude throughout the boundary layer. 

While many of the critical questions in heterogeneity driven flows can, and are, being addressed through carefully constructed LES, the field would benefit greatly from observational campaigns that include a measurement or indicator of heterogeneity driven flows. Observations can be challenging, because any observation would require three dimensional information of the atmospheric boundary layer to fully assess circulatory motions and internal boundary layers, and horizontal spatial coverage on the order of kilometers would be necessary for larger scale flows. A number of campaigns have sought to address the difficult task of exploring heterogeneity and land atmosphere interactions through dense networks of flux towers, often synthesized with modeling, other instrumentation and remote sensing, at a number of scales (mesoscale and sub-mesoscale) \cite{morrison_heat-flux_2022,lohou_model_2025,butterworth_connecting_2021,boone_land_2025,wang_how_2024}. However, studies have found that even with a relatively dense network of towers, secondary circulations can be a challenge to quantify directly and connect with surface heterogeneity without supplemental LES \cite{paleri_impact_2025}. Such networks are also expensive, and often limited to near surface characterization of any flow. Most observational campaigns that emphasize heterogeneous land atmosphere interactions instead focus on changes to surface fluxes, which, while influenced by heterogeneity driven flows, are affected by a large number of other processes and can't directly serve as an indicator of mesoscale heterogeneity driven flows \cite{boone_land_2025,lohou_model_2025,butterworth_connecting_2021}. 

In addition to observational systems, a strong metric quantifying the flows is needed. Any metric would need to be able to assess the phenomena in diverse atmospheric and surface conditions, while ideally being easily applicable in a gridded model context. Work in this area is increasingly essential, as development has already begun on model parameterizations for ESMs that account for heterogeneity driven flows. Attempts based only on modifying the land surface model component directly have seen limited success \cite{fowler_assessing_2024,huang_representing_2022}, and efforts are increasingly orientated towards modifying atmospheric boundary layer schemes. A model that divides the flow into two columns with a physics-based conceptual model for circulations has been developed \cite{waterman_twocolumn_2024}, as have simple models in the ocean context \cite{naumann_moist_2019}. Implementation is also ongoing in the Community Earth System Model (CESM and NASA GEOS \cite{arnold_representing_2024} ESMs for multi-plume eddy-diffusivity mass flux schemes that account for the one-dimensional atmospheric impact of surface heterogeneity, with potential for the future integration of two-dimensional or three-dimensional heterogeneity driven flows as well. For greater understanding and effective model parameterization of these phenomena, new methodology and metrics are necessary for direct or indirect observations of heterogeneity driven flows. 

In this work, we introduce a metric to serve as a potential indicator for heterogeneity driven flows, and examine the ability for a Doppler LiDAR based network to measure the impact of surface heterogeneity on the atmospheric boundary layer. 

Section \ref{sec:methods} describes the metrics used to evaluate spatial heterogeneity (section \ref{sec:methods_spatial}) and atmospheric impact (section \ref{sec:methods_dke}). Section \ref{sec:data} describes the details of the Doppler LiDAR network over the Atmospheric Radiation Measurement (ARM) site in the Southern Great Plains (SGP) and the satellite based measurements of GOES-LST used to evaluate spatial heterogeneity in this domain. In addition to these measurements, section \ref{sec:data} describes the LES experiments from \citeA{simon_heterogeneous_2024} that are used to contextualize and assess sensitivity in the observational results. Results are then presented for general characteristics of the measurement profiles (section \ref{sec:results_observ}), parameter sensitivity for screening the data (section \ref{sec:results_sensitivity}), the relationship between spatial heterogeneity and atmospheric impact (section \ref{sec:results_het}), and likely sensitivity of the observations to network design (section \ref{sec:results_network}).

\section{Methods} \label{sec:methods}
To assess the impact of the heterogeneity of surface heating on an overlying atmosphere, we need quantitative assessments of spatial heterogeneity and atmospheric impact that function well under idealistic and complex surfaces, as well as fully described atmospheres (in high resolution modeling) and partially described atmospheres (point based measurements).
\subsection{Describing Spatial Heterogeneity} \label{sec:methods_spatial} In idealized studies of surface heterogeneity for atmospheric impact, the surface is often arranged in a checkerboard pattern with patches of high and low surface temperature \cite{margairaz_surface_2020,lee_effect_2019,papangelis_saharan_2021}. Under these configurations, the key surface parameters are clear: the size of the patches and the difference in heating between them. For real (or realistic) land surfaces, neither of these parameters are immediately obvious, and as such we must employ similar analogues.
\par To assess the size of prevailing surface heating patterns, we use spatial covariance functions, a commonly applied method for understanding surface spatial statistics \cite{torresrojas_geostatisticsbased_2024,wikle_modern_2015,genton_cross-covariance_2015,zakeri_review_2021}. For convenience, temperature will be assumed to follow second-order weak stationarity, meaning that the covariance between any two points will only be a function of the separation between them. After computing the the covariance in surface temperature between points, and examining this covariance as a function of distance, the resulting curve can often be fit following an exponential form: 
\begin{equation} \label{eq:exp_fit}
    \rho(x) = A \exp\left(-\frac{x}{\lambda}\right), 
\end{equation}
where $A$ is a sample covariance and $\lambda$ is the correlation length. Generally, the correlation length will control how quickly the covariance between two points decreases as a function of separation distance. The correlation length parameter thus inherently captures both the general size of temperature patches as well as how they are organized among themselves. A larger correlation length generally implies larger, fewer, and more distinct patches of similar temperatures. The correlation length should then directly correlate with heterogeneity. Optimal values of the correlation length of heterogeneity tend to exceed 5-10 kilometers (\cite{lee_effect_2019, avissar_evaluation_1998,simon_heterogeneous_2024}) to best facilitate atmospheric impact in LES.  
\par Difference in the magnitude of heating also needs to be assessed. A larger difference in temperature between two patches will generally induce a stronger horizontal pressure gradient at the surface, leading to the possibility of stronger atmospheric response. The spatial standard deviation of the temperature field at the surface ($\sigma_{T_s}$), often normalized by the spatial mean of temperature to generate a coefficient of variation ($CV=\sigma_{T_s}/\overline{T}_s$, will be used to characterize the overall temperature variation. When examining only two patches (as in a checkerboard) $\sigma_T=0.5\Delta T$. Combining the correlation length and temperature variation into a single metric yields the following as the primary measure of thermal heterogeneity in this work:
\begin{align}
    h = \frac{\lambda_{T_s} \sigma_{T_s}}{\overline{T_s}}.
\end{align}
\par Whether the patterns of surface heating heterogeneity at the surface generate temperature and pressure differences in the atmosphere will depend partially on atmospheric conditions. As discussed in section \ref{sec:intro}, when the heterogeneity is oriented so that the surface heating gradient is parallel to a moderate or strong background wind, the surface heating pattern rarely persists into the atmosphere and impact on mesoscale flow is expected to be minimal. To capture impact of the relative orientation of the spatial heterogeneity to dominant flows, we leverage the angle between the surface heating gradient and large scale wind as $\alpha$.

\subsection{Quantifying Atmospheric Impact} \label{sec:methods_dke}
A method is also needed to assess the atmospheric impact. A number of recent studies have focused on measuring changes to dispersive fluxes, especially the dispersive heat flux, to examine the impact of surface heterogeneity on vertical heat transport in large areas \cite{morrison_heat-flux_2022,margairaz_surface_2020-1,akinlabi_dispersive_2022,papangelis_saharan_2021,paleri_impact_2025}. Others consider mesoscale fluxes that are very similar to dispersive fluxes and functionally the same if Taylor's frozen turbulence hypothesis is accepted \cite{maronga_large-eddy_2013}. These metrics, however, are specific to a given scalar and, due to a strong dependence on downdraft and updraft frequency which will vary spatially, may face large uncertainties when measured at only a few locations in an observational network as opposed to a full model grid. Analysis focusing on regional, dispersive flow statistics may still be a useful route for examining heterogeneity driven impact.

Consider the three-dimensional flow field for the atmosphere over a region, with wind components $u$, $v$, and $w$ in the three principal coordinate directions. A typical way of accounting for turbulence in the ABL is performing a Reynolds' decomposition, whereby a flow variable, $f$, is split into a time-averaged component and another component representing turbulent fluctuations in time. Such a variable can be viewed as the result of turbulent perturbations ($f'$) around a mean value ($\overline{f}$):
\begin{equation}
    f = \overline{f} + f'
\end{equation}
In a setting where flow variables vary temporally due to turbulence and spatially due to the effects of heterogeneity, a slightly different and more complete averaging scheme can be used. This scheme incorporates a time average, followed by a spatial average large enough to eliminate fluctuations on the scale of local turbulence (\cite[p. 80]{raupach_averaging_1982}). Such a scheme can be understood through a triple decomposition of $f$ akin to the Reynolds' decomposition:
\begin{align}
    f = \underbrace{\overline{\langle f \rangle}}_{\substack{\text{Spatio-temporal}\\\text{Average}}} + \underbrace{\overline{f}''}_{\substack{\text{Spatial Perturbation}\\\text{of Time Average}}} + \underbrace{f'}_{\substack{\text{Local Temporal}\\\text{Perturbation}}}.
\end{align}

The mean value of the flow variable is now averaged over space and time while the local temporal perturbation, $f'$, remains the same. The middle term in the decomposition captures the spatial variations in the time average of the variable (\cite[p. 80]{raupach_averaging_1982}). Therefore, this variable decomposition isolates the effects of local turbulence from variations arising from spatial heterogeneity. As mentioned in \citeA{raupach_averaging_1982}, the spatial average must be large enough to remove local turbulent effects. Applying the above averaging scheme to the kinetic energy of the flow field yields the following:
\begin{equation} \label{eq:mke_dke_tke}
    \frac{1}{2} \langle u_i u_i\rangle =  \underbrace{\frac{1}{2}\langle \overline{u}_i \rangle \langle \overline{u}_i \rangle}_\text{MKE}  + \underbrace{\frac{1}{2}\langle \overline{u}''_i \overline{u}''_i \rangle}_\text{DKE} + \underbrace{\frac{1}{2}\overline{\langle u'_i u'_i\rangle}}_\text{TKE}\ 
\end{equation} where $i$ represents iteration over the three cartesian coordinates $x$, $y$, $z$. 

The first term represents the kinetic energy arising from mean flow over a given area (mean kinetic energy, or MKE), while the third term represents the kinetic energy from turbulent fluctuations (turbulent kinetic energy (TKE)). The middle term represents the kinetic energy arising from time-averaged spatial variations in the velocity field, or Dispersive Kinetic Energy (DKE) (\cite[p. 86]{raupach_averaging_1982}). Previous work from \citeA{waterman_surface_2025} has shown a clear relationship between surface heterogeneity and a metric closely related to DKE. In \citeA{waterman_surface_2025}, they further assume Taylor's frozen turbulence hypothesis to replace the temporal averaging with a (small scale) spatial averaging and term this Mesoscale Kinetic Energy (MsKE) which should be analogous to DKE at these scales and useful in Numerical Weather Prediction where grid cells do not explicitly resolve turbulence. 

As in \citeA{waterman_surface_2025}, the DKE can be vertically integrated through the atmosphere and weighted by air density:
\begin{equation}
    DKE_v =\frac{1}{\overline{\rho}z_{top}} \int_0^{z_{top}} \rho(z)\  \left[\langle \overline{u}''\overline{u}'' \rangle + \langle \overline{v}''\overline{v}'' \rangle + \langle \overline{w}''\overline{w}'' \rangle \right]\  dz
\end{equation} where $z_{top}$ is the integration limit and $\rho$ is the air density.
Each term within the parentheses represents the spatial variance of each time-averaged velocity component, which will be computed over the region or set of points from which the velocity is being measured. Discretizing the atmosphere into layers of varying thickness, we arrive at the following equation to calculate DKE for any region with known vertical wind profiles: 
\begin{equation} \label{eq:int_DKE}
    DKE_v = \frac{1}{2}\frac{1}{\rho z_{top}} \sum_{k=0}^{z_{top}} \rho_k (\sigma_{\overline{u}_k}^2 + \sigma_{\overline{v}_k}^2 + \sigma_{\overline{w}_k}^2) \Delta z_k
\end{equation} where the overline indicates temporal averaging (but not spatial averaging).
We will further be interested in DKE as a fraction of MKE.  While the goal is to isolate kinetic energy that we can attribute to local, mesoscale, heterogeneity driven flows, signals from larger scale atmospheric motion may still be picked up. From both a theoretical and observational standpoint, it is still possible that effects from the mean flow are picked up in the signal of DKE. Normalization of DKE by MKE may more accurately assess the relative strength of dispersive terms compared to the rest of the flow. Since we cannot fully partition the $DKE$ caused by local heterogeneity and $DKE$ initiated by the mean flow, analysis of both the ratio which we define as DKE fraction $DKE_f=\frac{DKE}{MKE}$ and $DKE$ in isolation are important. If the contribution from the mean flow is small, $DKE$ would be a more appropriate tool to assess heterogeneity driven flows and the ratio would become highly sensitive to background velocity. This is particularly problematic when $\alpha$ approaches $90^{\circ}$ where one would expect similar heterogeneity driven flows regardless of the wind velocity and MKE. By contrast, if the contribution to DKE from MKE is larger the ratio becomes more pertinent; as such, both are presented throughout the work. The DKE ratio is also useful for non-dimensional analysis. Another pertinent non-dimensionalization may be DKE/TKE, however the LiDAR product is insufficient to assess TKE and as such this is not explored as part of this work. It is also notable that the passing of large frontal systems or convective storms would also be expected to show up significantly in $DKE$, which later analysis attempts to address. The MKE can also be vertically integrated through the atmosphere:
\begin{equation} \label{eq:int_MKE}
    MKE_v = \frac{1}{2}\frac{1}{\overline{\rho}z_{top}} \sum_{k=0}^{z_{top}} \rho_k (\langle\overline{u}\rangle^2 + \langle\overline{v}\rangle^2 + \langle\overline{w}\rangle^2) \Delta z_k. 
\end{equation} where the overline and angle bracket together indicate spatial and temporal averaging.
Both $DKE_v$ and $MKE_v$ are calculated at hourly time-steps and then time averaged across a given day from 10am to 4pm local time. The vertical integration is conducted from zero to 500 meters to highlight near surface response where stronger coupling between the land and atmosphere is expected. Much of the analysis appears insensitive to exact specification of $z_{top}$. Specifics for how the $DKE_v$ and $MKE_v$ are calculated for observational data is discussed in section \ref{sec:data_lidar}. 

\section{Data and Processing}\label{sec:data}
The analysis in this manuscript is performed at the U.S. Department of Energy's Atmospheric Radiation Measurement (ARM) user facility in the Southern Great Plains (SGP). There is a wide variety of instrumentation and site networks within the ARM-SGP facility, however this research focuses on leveraging satellite remote sensing from GOES (section \ref{sec:data_goes}), a series of LES simulations over ARM-SGP from \citeA{simon_heterogeneous_2024} (section \ref{sec:data_les}), and ground based light detection and ranging sites (LiDAR) (section \ref{sec:data_lidar}).
\subsection{GOES Land Surface Temperature} \label{sec:data_goes} 
The National Oceanic and Atmospheric Administration's (NOAA) Geostationary Operational Environmental Satellites (GOES) are the main operational, geostationary weather satellites in the western hemisphere \cite{desai_multi-sensor_2021}. Retrievals from the new generation GOES-16 and GOES-17 satellites enables a land surface temperature (LST, $T_s$) product at hourly temporal resolution and approximately 2 km spatial resolution at nadir. While the LST product is of high quality, it is sensitive to cloud coverage as the primary retrieval bands for the LST product are non-cloud penetrating \cite{desai_multi-sensor_2021}. As such, clear sky conditions are needed for the analysis present in this work (defined as $>90\%$ of LST values in the domain available). We further assume that the morning spatial patterns of temperature persist throughout the day and ultimately set the atmospheric response. This should be a good assumption when significant heterogeneity exists that is caused by long-term persisting variables (i.e. soil moisture, topography, vegetation), however spatio-temporal persistence of LST fields are notably complex \cite{torresrojas_geostatisticsbased_2024}. As a result of this assumption, the LST fields from GOES used for this work are from 9 and 10am local time. This choice also allows for afternoon cloud development (which heterogeneity driven flows are known to cause in LES \cite{lee_effect_2019,simon_heterogeneous_2024}) while still maintaining the clear sky requirement in the GOES data. Prior to calculation of the metrics described in section \ref{sec:methods_spatial}, a weak gaussian filter is applied to the GOES LST field to reduce the impact of the smallest scale (2 km resolution) spatial perturbations which are less likely to be relevant for the heterogeneity driven flows of interest and instead add spatial noise.

\subsection{Doppler LiDAR} \label{sec:data_lidar} 
\begin{figure}
\noindent\includegraphics[width=\textwidth]{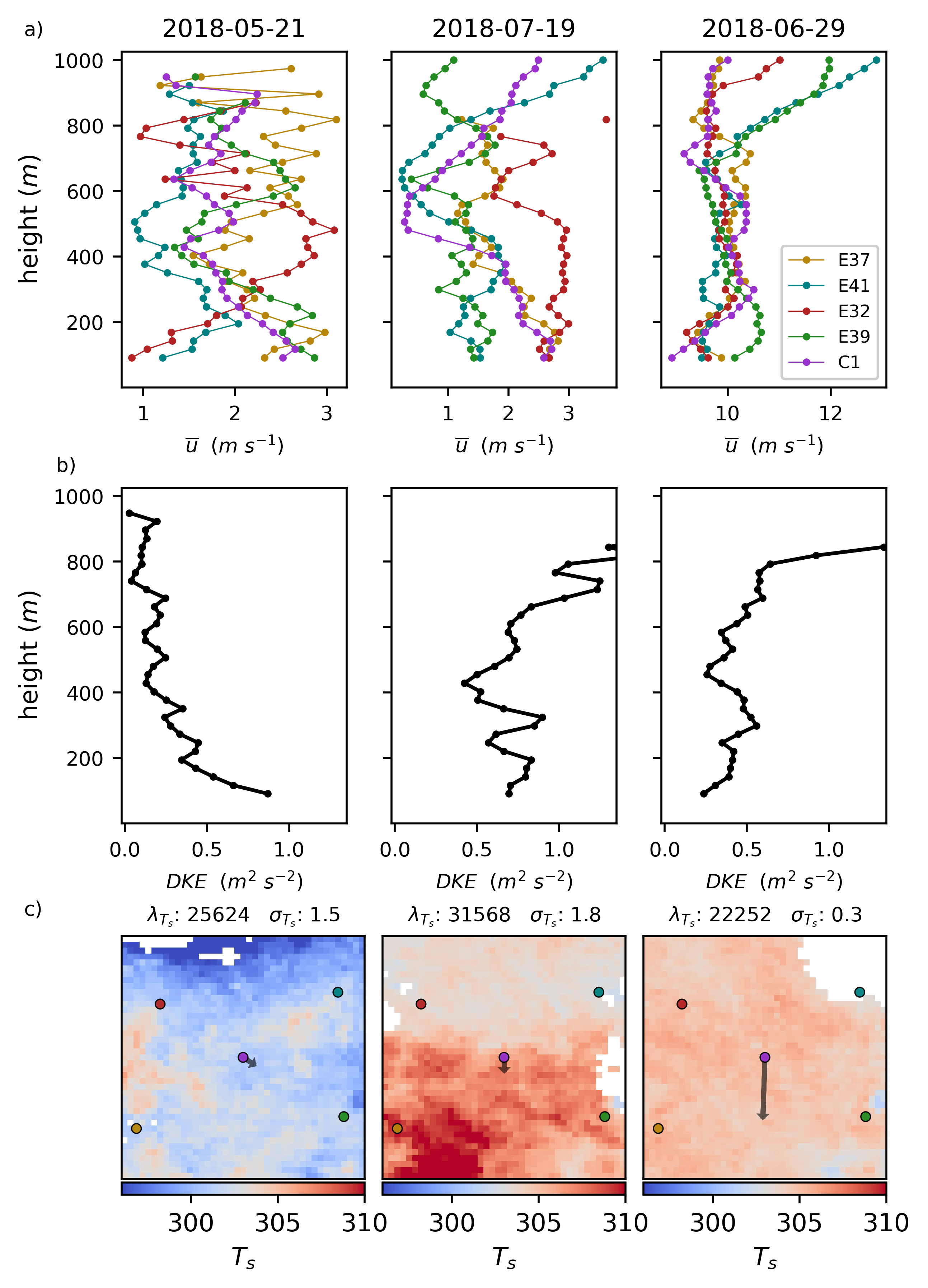}
    \caption{Vertical profiles for three example days (2018-05-21, 2018-07-19 and 2018-06-29) of one-hour averaged horizontal velocity at noon local time \textbf{(a)} and DKE from those profiles \textbf{(b)}. Profiles of velocity are colored according to the site in the LiDAR network they come from. Also shown are the GOES LST surfaces \textbf{(c)} from each day with the location of the LiDAR measurements shown. Mean wind direction is also plotted as an arrow from the center site. For the surfaces, the lengthscale of heterogeneity and standard deviation of land surface temperature are also shown.}
    \label{fig:example_prof}
\end{figure}

Doppler light detection and ranging (LiDAR) networks have a variety of useful characteristics for detecting hetergoeneity driven flows and computing DKE. Doppler LiDAR can provide multiple kilometer profiles of the three components of velocity and, depending on the scanning mode, can produce fairly high temporal resolution data in contrast to traditional radiosonde based atmospheric profiles. Doppler LiDAR networks are also starting to be used more widely to measure characteristics of the boundary layer, with a significant recent push to expand such networks \cite{wulfmeyer_new_2018} and others recently established \cite{hohenegger_fesstval_2023}. The SGP-ARM user facility has many remote sensing instruments available, including Doppler LiDAR profilers. The Doppler LiDAR network at ARM-SGP consists of five locations, including the central observatory C1-Lamont located in the middle of four auxiliary sites, E32, E41, E37 and E39, which are configured in a rectangle with each auxiliary site 56 to 77 km away from each other around C1. Figure \ref{fig:example_prof}c shows the configuration of these five sites on top of the GOES LST field for three days. The five sites use Doppler LiDAR Halo Photonics Stream Line scanning systems, and lead to products of the three velocity components as well as wind direction at 10 minute temporal resolution and approximately 26 m vertical resolution. The LiDAR instrumentation cannot effectively measure close to the surface, and the first measurement is at 91 meters above ground level and extends several kilometers into the atmosphere. These products are quality assured, corrected and provided by ARM, with additional details on their deployment and the Doppler LiDAR methodology found in \citeA{newsom_doppler_2022}. Meteorological information is also provided at a singular height for each site; if the meteorological values for wind speed or wind direction deviate significantly ($>45^{\circ}$) at this height, the data is discarded. Figure \ref{fig:example_prof}a shows an example of the profiles of horizontal velocity from the LiDAR measurements.
\subsubsection{Calculating DKE from LiDAR} \label{sec:data_lidar_dke}
To calculate DKE from a network of vertical profiles, as in figure \ref{fig:example_prof}, the velocities are first temporally upscaled to hourly profiles. At each vertical level in the atmosphere, the sample variance of the five measurements is computed for each velocity component ($u$,$v$, $w$). This sample variance captures the spatial deviations from a temporally averaged velocity, and therefore the sum of these variances together is the DKE for that level as in equation \eqref{eq:int_DKE}. The profile of the DKE for three example days is shown in figure \ref{fig:example_prof}. To calculate a single DKE value for each hour, this profile is integrated as in equation \eqref{eq:int_DKE} on an hourly basis to a height of 500 m. The integration is done over this relatively low height to prioritize the surface influence on the atmosphere and constrain the calculation to the boundary layer for the majority of daytime hours. The hourly $DKE_v$ is then averaged across the day from 10:00 to 4:00 local time, discounting early hours where the boundary layer is developing and later hours where it may be beginning to collapse.

\subsection{Large Eddy Simulations} \label{sec:data_les} 
While the focus of this work is primarily on LiDAR measurements, a set of Large Eddy Simulation (LES) experiments are used to contextualize the results and examine the uncertainty from the configuration of the LiDAR networks. The LES runs employ a modified WRF-LES following the configuration in \cite{simon_semicoupling_2021} and \cite{simon_heterogeneous_2024}. Additional details can be found in these publications, which describe the configuration \cite{simon_semicoupling_2021} and summarize results of the extended simulations \cite{simon_heterogeneous_2024}. We will summarize the key points here. In total, 92 single day summer simulations are conducted at 250m horizontal and 30m vertical resolution on a 130 km by 130 km by 12 km domain. The domain is based around the ARM-SGP observatory, like the LiDAR network, and uses data from their measurement network is used to set initial meteorological conditions. The simulations are conducted with two different surface configurations; one is heterogeneous using land surface model (LSM) output to produce a realistic surface at the same resolution as the LES. The other has a homogeneous surface with sensible and latent heat fluxes forced to be the same domain average as the heterogeneous case. The model is run with periodic boundary conditions. The 92 days are all shallow convection days. DKE can be calculated for LES as described in section \ref{sec:methods_dke} for the whole domain. To assess uncertainty in the DKE calculations from LiDAR, we generate virtual LiDAR networks for various numbers of towers; for each network size, 100 virtual networks are made by randomly selecting gridcells from the LES. DKE is then calculated for these virtual networks using the methods described in section \ref{sec:data_lidar_dke}.

\section{Results} \label{sec:results}
\subsection{Observing Heterogeneity Driven Flows} \label{sec:results_observ}
\begin{figure}
\noindent\includegraphics[width=\textwidth]{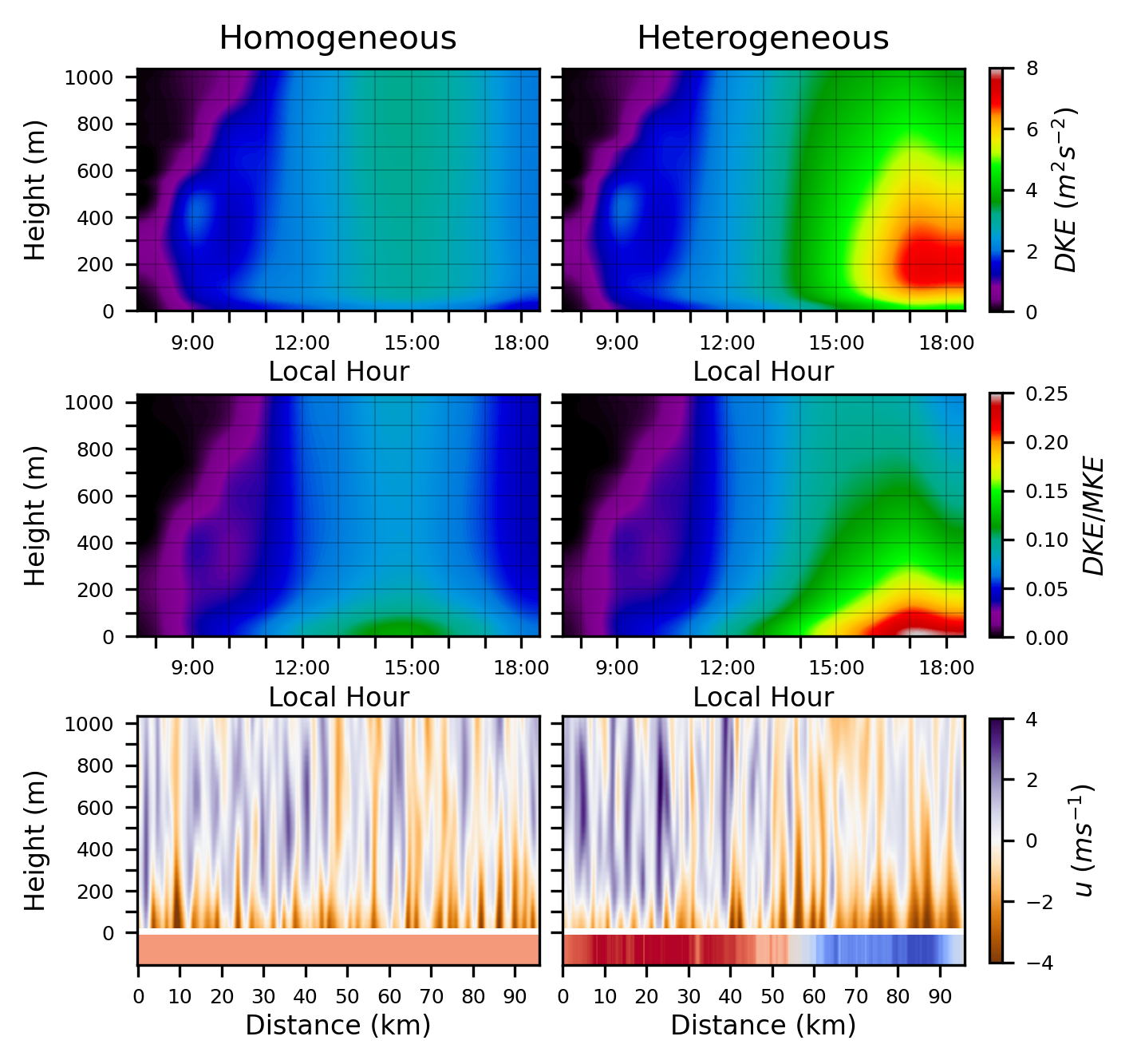}
    \caption{LES profiles of DKE (top) and DKE/MKE (middle) through time in a large eddy simulation from 2016-06-25 as well as a horizontal cross section of the horizontal velocity and the surface temperature (bottom). The profiles on the left are from a homogeneous surface LES simulation where heterogeneity driven flows cannot be developed, and the profiles on the right are from a heterogeneous surface LES simulation.}
    \label{fig:les_prof}
\end{figure}

Heterogeneity driven flows over land have rarely been observed directly and quantitatively as discussed in section \ref{sec:intro}. In LES, heterogeneity driven flows, and specifically circulations driven by heterogeneity, are readily apparent. Previous work using the same LES experiments described in section \ref{sec:data_les} clearly show circulatory motion driven by heterogeneity \cite{waterman_twocolumn_2024,simon_heterogeneous_2024}. In figure \ref{fig:les_prof}, we show the profiles of DKE, DKE/MKE, horizontal velocity and surface heating for an LES day previously explored in detail in \citeA{waterman_twocolumn_2024}. In the bottom panel, we see evidence of breeze formation over the heterogeneous surface; at the boundaries between warm and cool surface in the bottom panel (around 0 km, 50km and 100km since the boundary conditions are periodic) flow is predominately following the temperature (and pressure/density) gradient from cool to warm. By comparison, the homogeneous flow does not show consistent horizontal trends in velocity. The differences in flow cause differences in the DKE and DKE/MKE profiles between the homogeneous and heterogeneous cases as well. While profiles are nearly identical between the homogeneous and heterogeneous case through about noon local time, the afternoon profiles diverge as both the DKE ratio and the DKE increase significantly in the heterogeneous case while there is only a mild increase in the homogeneous case. Additionally, both quantities continue to grow and persist into the late afternoon/early evening in the heterogeneous case whereas they peak around 14:00 or 15:00 in the homogeneous case. This does not directly indicate that the DKE or the DKE ratio is directly measuring only these heterogeneity driven breezes and circulations; if this were the case we would expect no significant values or diurnal cycle in the homogeneous case. This does, however, indicate that these breezes are likely captured within the DKE and the DKE ratio as significant components of the measurements, up to 75\% of the values if we do a simple subtraction of the homogeneous DKE profiles from the heterogeneous. 

In the LiDAR measurements, with only five or less points in the network, it is much more difficult to identify breeze structures specifically (as opposed to other heterogeneity driven processes impacting the atmosphere) with confidence. This is part of the benefit of DKE as a more process neutral indicator for potential heterogeneity driven flow impacts. Regardless, there is some indication in the data that circulations or breezes may exist and DKE may serve as an indicator for them. In figure \ref{fig:example_prof}, on 2018-05-21 the profiles of horizontal velocity form two distinct clusters based on velocity, with E32 (red) and E41 (teal) both on the north side of the domain at the boundary between the cool and warm area and C1 (pink) E37 (yellow) and E39 (green) all on the southern side of the domain in a warmer area. There is similar clustering in 2018-07-19, with the spatial organization by velocity primarily east west in this case. The organization is also perpendicular to the dominant direction of the velocity, which is the organization one would expect for circulations. The spatial organization of velocity, especially relative to the background velocity and the surface heating patterns, may be indicators of breezes. 2018-06-29 has very little surface heterogeneity as well as strong background winds; this difference between the three days is reflected in the near surface DKE profile (first few hundred meters), which is larger in both of the more heterogeneous days.

\begin{figure}
\noindent\includegraphics[width=\textwidth]{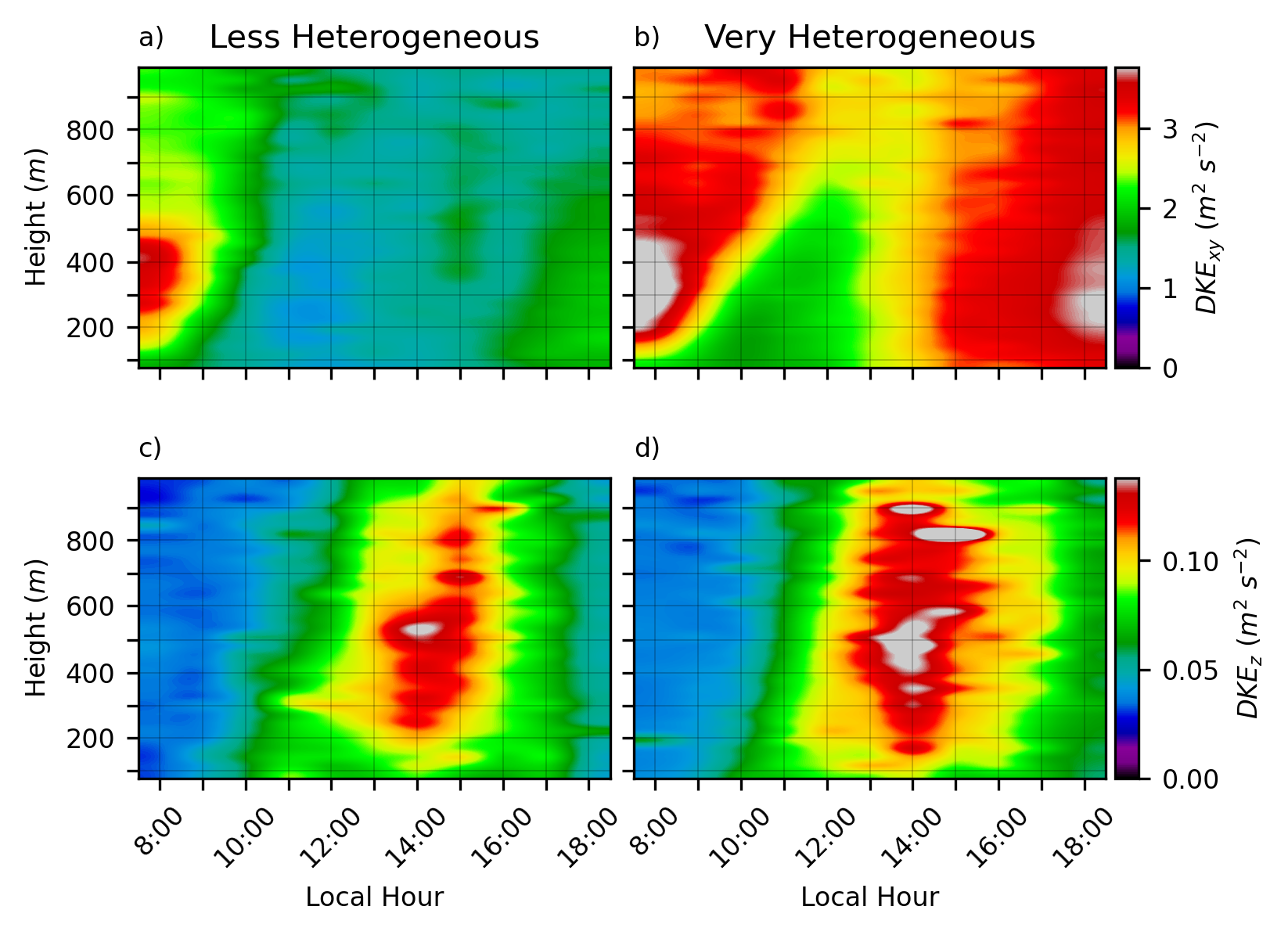}
    \caption{LiDAR profiles through time of the horizontal component of DKE (a,b) and the vertical component of DKE (c,d) averaged across a group of more homogeneous days (a,c) and more heterogeneous days (b,d). More heterogeneous days are all LiDAR days for which $\sigma_{T_s}>0.75$ and less heterogeneous days are those with $\sigma_{T_s}<0.75$}
    \label{fig:dke_prof}
\end{figure}

We can also examine the DKE profiles through time in LiDAR. Figure \ref{fig:dke_prof} breaks down the diurnal cycle of the DKE profile averaged across a group of days with less heterogeneous surface heating (figure \ref{fig:dke_prof}a,c) and more heterogeneous surface heating (figure \ref{fig:dke_prof}b,d) as well as horizontal and vertical components. The DKE is substantially larger for the heterogeneous days than the homogeneous days, nearly double, indicating a clear influence from surface heterogeneity on DKE and its potential to measure the impact of heterogeneity on the atmosphere. The breakdown into vertical $DKE_z$ and horizontal $DKE_{xy}$ indicates that horizontal motion dominates the metric. During daytime, the vertical component is at most 10\% of the total DKE, and can be as low as 1 or 2\%. This was also seen in the LES experiments (not shown) and in continental scale NWP \cite{waterman_surface_2025}. While both horizontal and vertical components of DKE appear to be larger during the heterogeneous days, the figure also indicates that $DKE_{xy}$ may be be more sensitive to heterogeneity than $DKE_{z}$. The lack of a strong vertical dependence for $DKE_z$ also indicates that the vertical component may not be absorbing strong signals from the surface in contrast to the horizontal component. The vertical component of DKE is essentially the spatial vertical velocity variance, and should be closely related to the strength and frequency of updrafts and downdrafts in the region. Figure \ref{fig:dke_prof}c,d do seem to indicate that larger vertical velocity variances push deeper into the boundary layer, possibly suggesting stronger deeper updrafts are slightly more prevalent. Deeper updrafts are associated with convective initiation and increased dispersive fluxes, both previously reported in LES simulations of heterogeneity driven flows. While the vertical DKE increase is small, the metric does not illustrate the relative spatial organization of updrafts and downdrafts. If the updrafts are more aligned with warmer regions and downdrafts with cooler in more heterogeneous terrain, the dispersive fluxes could be significantly larger despite little to no change in the vertical DKE or vertical velocity variance, as found in earlier LES experiments.

In the morning, $DKE_{xy}$ is large outside of the convective boundary layer, in contrast to $DKE_{z}$. This large $DKE_{xy}$ value in an atmospheric layer that is supposed to be largely decoupled from the local surface due to the strong stability is surprising if DKE is thought to only capture heterogeneity driven flow impacts. However, it is relatively unsurprising if other atmospheric phenomena that could drive DKE, such as a strong low level jet that does not cover all the LiDAR locations, are considered. Under a strong low level jet, we would also expect MKE to be large, but if both DKE and MKE are large the ratio between them would presumably be unaffected. 

\begin{figure}
\noindent\includegraphics[width=\textwidth]{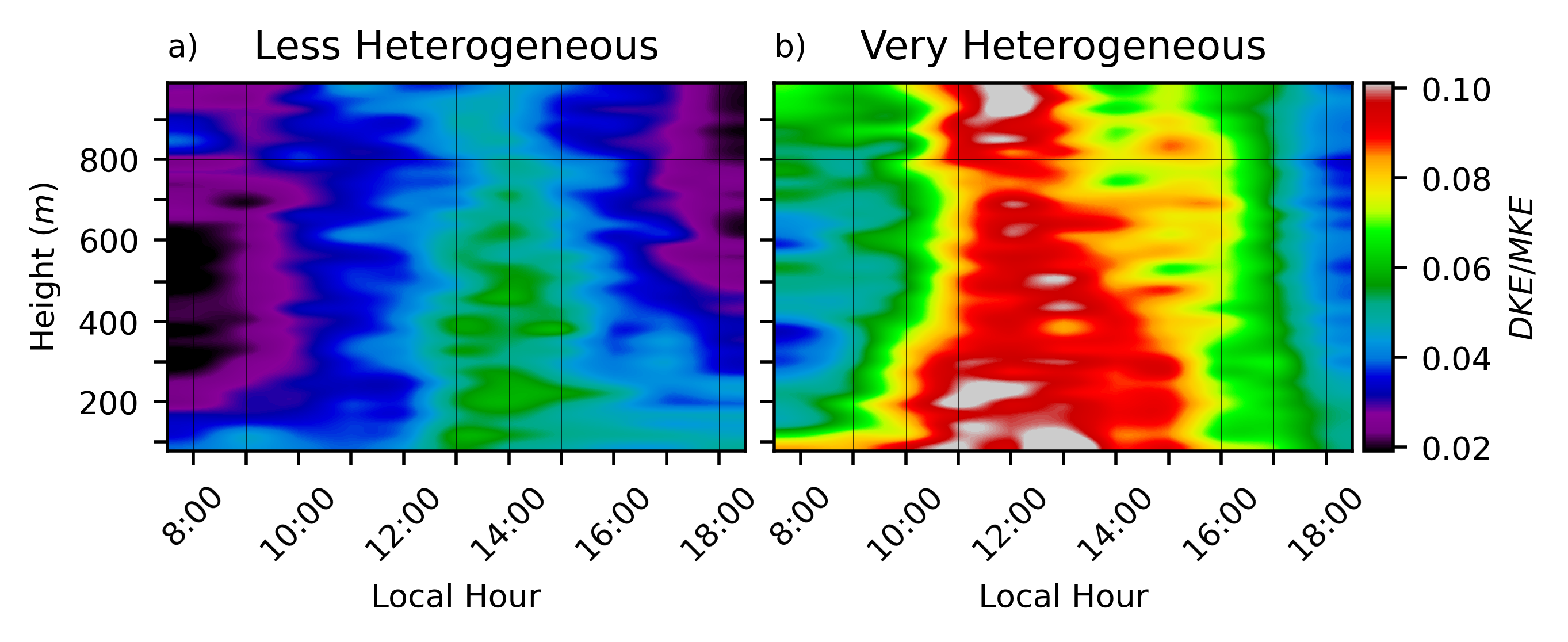}
    \caption{LiDAR profiles through time of DKE/MKE averaged across a group of more homogeneous days (a) and more heterogeneous days (b). More heterogeneous days are all days for which $\sigma_{T_s}>0.75$ and less heterogeneous days are those with $\sigma_{T_s}<0.75$}
    \label{fig:ratio_prof}
\end{figure}

The high DKE values in the Stable Boundary Layer are no longer present when examining the DKE ratio in figure \ref{fig:ratio_prof}. As with the DKE profile, there is a strong increase over the more heterogeneous days when compared to the less heterogeneous days. Unlike the DKE profile, there is a strong diurnal cycle peaking around 12:00 or 13:00 and then decaying throughout the afternoon. 

The DKE and DKE ratio profiles from LiDAR (figures \ref{fig:dke_prof},\ref{fig:ratio_prof}) show some significant differences from the LES profiles (figure \ref{fig:les_prof}). While the LES and LiDAR are not directly comparable (different days/collection of days) the general characteristics of the profiles can be compared. The DKE grows more slowly in magnitude throughout the day in the LES, however the vertical extent of the surface-influenced DKE grows more rapidly with less vertical variation in the profile in the morning. The high DKE stable boundary layer does not exist in the LES. LES has well known issues in representing the stable boundary layer, particularly at such a coarse resolution during a transition period and with little spin up time to develop \cite{couvreux_intercomparison_2020}. Apart from the stable boundary layer, and the slower vertical growth, the profiles are ultimately quite similar between LES and LiDAR. When examining the DKE ratio, however, there are more apparent differences. The LES has a much later peak in DKE ratio compared to LiDAR and in general persists later into the day; this may be due to the periodic boundary conditions in the LES, which will encourage longer persistence in atmospheric behavior and allow for a stronger atmospheric gradient in heating than would exist with non-periodic boundaries. LES, additionally, will not see the impact from the previous days heterogeneity which may be affecting the results above the convective boundary layer in the morning. 

These results show that the LiDAR profiles are generally analogous to well established LES behavior, with a few exceptions. The DKE ratio, in particular doesn't match perfectly with what would be expected from LES and indicate that future exploration of heterogeneity driven flows in LES would benefit greatly from more realistic boundary conditions and a properly developed stable boundary layer. In both LES and LiDAR, there is both larger DKE and DKE ratio when the surface heating is more heterogeneous, representing a clear impact of the surface on the flow in the convective boundary layer.

\subsection{Conditions for Heterogeneous Land-Atmosphere Coupling} \label{sec:results_sensitivity}
As discussed in greater detail in section \ref{sec:intro}, there are a number of factors that we would expect to modulate the coupling between the land and atmosphere or affect DKE independent of local properties, even if the patterns of surface heating are pronounced. Three factors in particular are well established in the literature: (1) the magnitude of the background wind velocity, (2) the orientation of the wind relative to the surface heterogeneity and (3) the influence of non-local flows including jets, flows driven by large (100km+) scale heterogeneity in heating or topography, frontal systems, advection of upstream heterogeneity impacts and deep convection. To examine when the coupling between surface heterogeneity and DKE is strongest, a number of filters are explored relating to the three factors above. Clear sky conditions are required in the morning to retrieve the LST fields, which will remove some convective storm conditions. Data may also be filtered for any daytime precipitation to remove days where afternoon storms (deep convection) develop. It is notable, however, that heterogeneity driven circulations are expected to impact convective initiation and removing those days may be unwise. The wind velocity at 1 km, $u_g$, is also used as a filtering variable. In an attempt to further reduce the impacts of frontal boundaries, and convergence or divergence associated with deep convection, the relative vorticity is used: 
\begin{equation}
    |\zeta|=\frac{\partial v}{\partial x}-\frac{\partial u}{\partial y}.
\end{equation}
Relative vorticity is expected to be large in those conditions. However, as with precipitation, care must be taken in the selection of a relative vorticity filter as heterogeneity driven circulations would also likely display significant vorticity. Background winds from the High-Resolution Rapid Refresh (HRRR) atmospheric model fields will be used to compute relative vorticity over the region \cite{dowell_high-resolution_2022}. Additionally, the angle between the background wind velocity and the direction of the mean gradient of the LST field, $\alpha$, is computed as described in section \ref{sec:methods_spatial} and used as a screening parameter.

\begin{figure}
\noindent\includegraphics[width=3in]{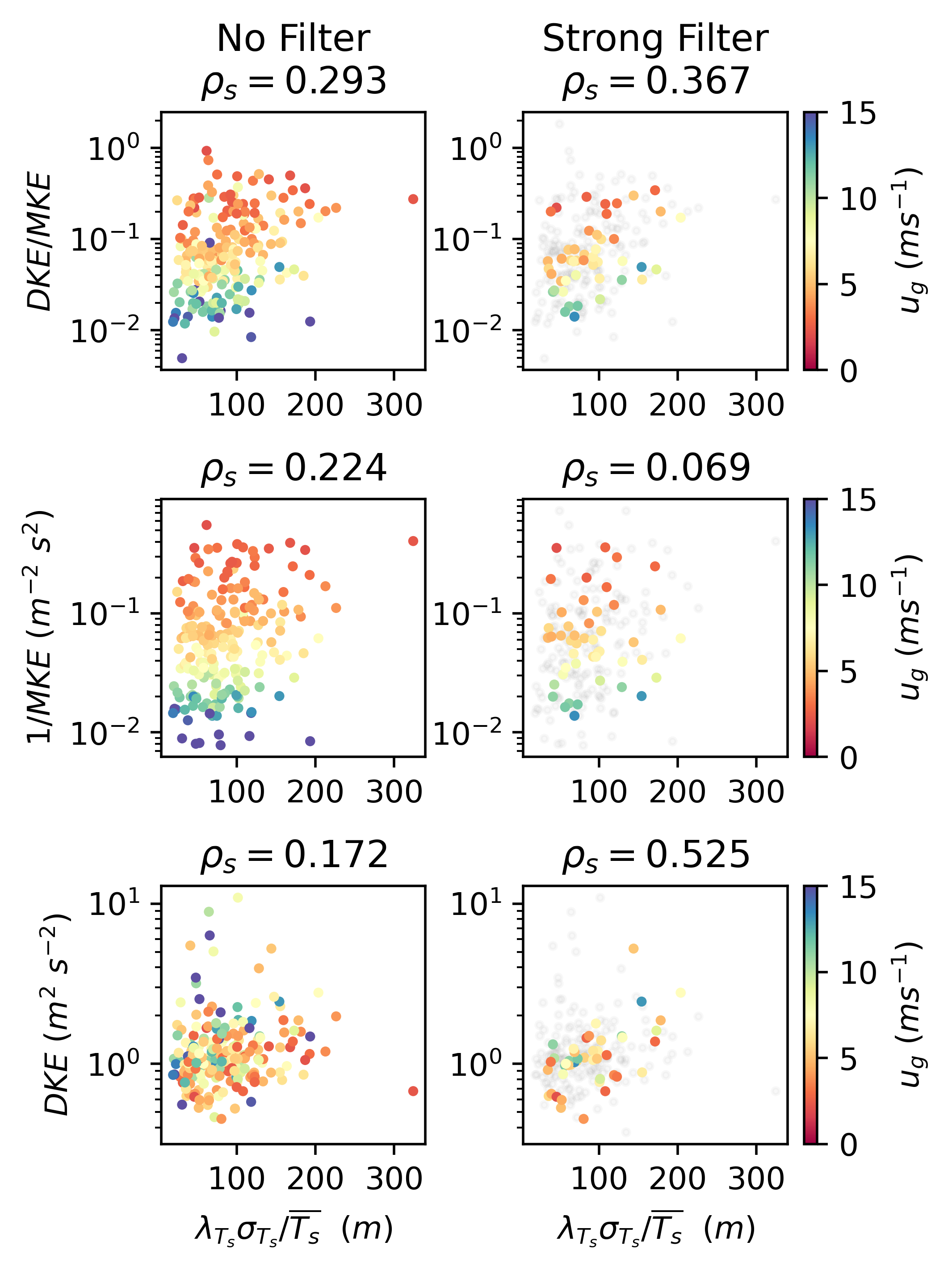}
    \caption{Scatterplots illustrating the relationship between the heterogeneity parameter $\lambda_{T_s}\sigma_{T_s}/\overline{T}_s$ and DKE/MKE (top), 1/MKE (middle) and DKE (bottom). The relationship is shown for all LiDAR days (left) and for a selection after a strong filter is applied (right). Spearman rank correlation coeficient is also shown. Scatter is colored according to the wind velocity at $1$ km ($u_g$). The strong filter restricts the analysis to days with at least 3 sites reporting, $u_g<15\ ms^{-1}$, $\alpha\geq70^{\circ}$, no daytime precipitation, and weak large-scale forcing $|\zeta|<2.5\times10^{-4}$}
    \label{fig:filter_1}
\end{figure}

Figure \ref{fig:filter_1} shows the heterogeneity parameter compared to the DKE, DKE ratio, and 1/MKE for each day of analysis, totaling 273 days before any screening is applied aside from clear sky mornings. On the left column of figure \ref{fig:filter_1}, with no filtering, there is a weak correlation between the heterogeneity parameter and the three variables, with the correlation strongest for the DKE ratio and weakest for the DKE. The correlation between the heterogeneity parameter and 1/MKE is surprisingly high and it is clear from the figure that at least some of the DKE ratio relationship with surface heterogeneity is somehow being driven by the background wind. In the right column of figure \ref{fig:filter_1}, a filter is applied to isolate for days where atmospheric conditions are expected to be more amenable to heterogeneity driven flows as discussed above, restricting the analysis to days with at least 3 sites reporting, $u_g<15ms^{-1}$, $\alpha\geq70^{\circ}$, no daytime precipitation, and weak large-scale relative vorticity $|\zeta|<2.5\times10^{-4}$. Once this filter is applied, the correlation with the heterogeneity parameter increases slightly for DKE ratio, decreases significantly for 1/MKE, and increases significantly to 0.525 for DKE. 

\begin{figure}
\noindent\includegraphics[width=\textwidth]{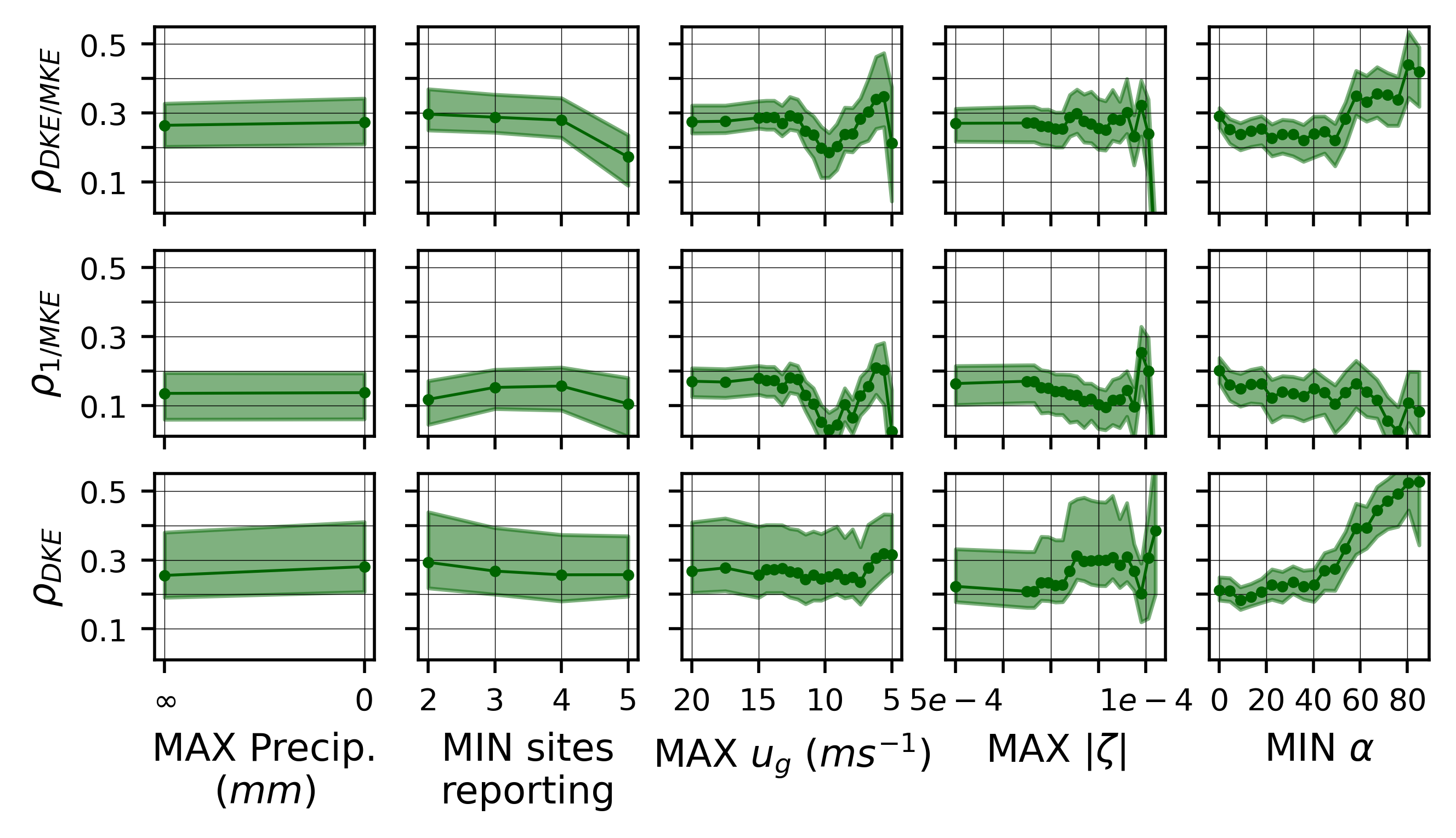}
    \caption{The median correlation coefficients between the heterogeneity parameter $\lambda_{T_s}\sigma_{T_s}/\overline{T}_s$ and DKE/MKE (top), 1/MKE (middle) and DKE (bottom), and the inter-quartile range of the coefficients, are shown across the filter parameter space. Five filters are explored; maximum precipitation, sites reporting, maximum $u_g$, max relative vorticity $|\zeta|$, and minimum angle between the wind and temperature gradient $\alpha$. Each point on each subplot shows the median correlation coefficient and inter-quartile range for all parameter sets with the given restriction. In all subplots, filters go from less restrictive (left) to more restrictive (right).}
    \label{fig:filter_2}
\end{figure}

The selection of exact filter cutoff values is somewhat arbitrary, so a more comprehensive analysis is necessary to make any strong conclusions. The full filter cutoff parameter space is explored in figure \ref{fig:filter_2}, including a total of 64,000 screening parameter combinations. The most consistently important filter is the angle between the surface gradient and the background wind, $\alpha$. Restricting to days where $\alpha$ is high (wind and heterogeneity run perpendicular to each other), greatly increases the correlation both the DKE ratio and DKE have with the heterogeneity parameters while the correlation with 1/MKE collapses. This makes sense; heterogeneity driven flows are expected to be strong when background wind is not overwhelming the surface heating pattern. MKE is very strongly related to background wind speed, which should matter less if it is perpendicular to the heating gradient. Relationships with the other filter values are less clear. Filtering high wind velocity has a mixed affect on the DKE ratio and 1/MKE. When velocities between $10$ and $15\ ms^{-1}$ are removed the correlation decreases and then recovers, in part because days in this range all have low values for the heterogeneity parameter, buoying the correlation (see the blue points in figure \ref{fig:filter_1}) so when they are removed the correlation drops. The DKE becomes more correlated with heterogeneity under very strong velocity filters. If DKE is capturing heterogeneity driven circulations, this would be consistent with literature that suggests even moderate wind velocities may substantially weaken these flows. The impact of relative vorticity is inconsistent when the filter is strong, possibly due to low sample sizes; a moderate filter, however, appears to decrease the correlation for MKE while increasing the correlation for DKE. 

\subsection{Metrics for Land Heterogeneity} \label{sec:results_het}
\begin{figure}
\noindent\includegraphics[width=5in]{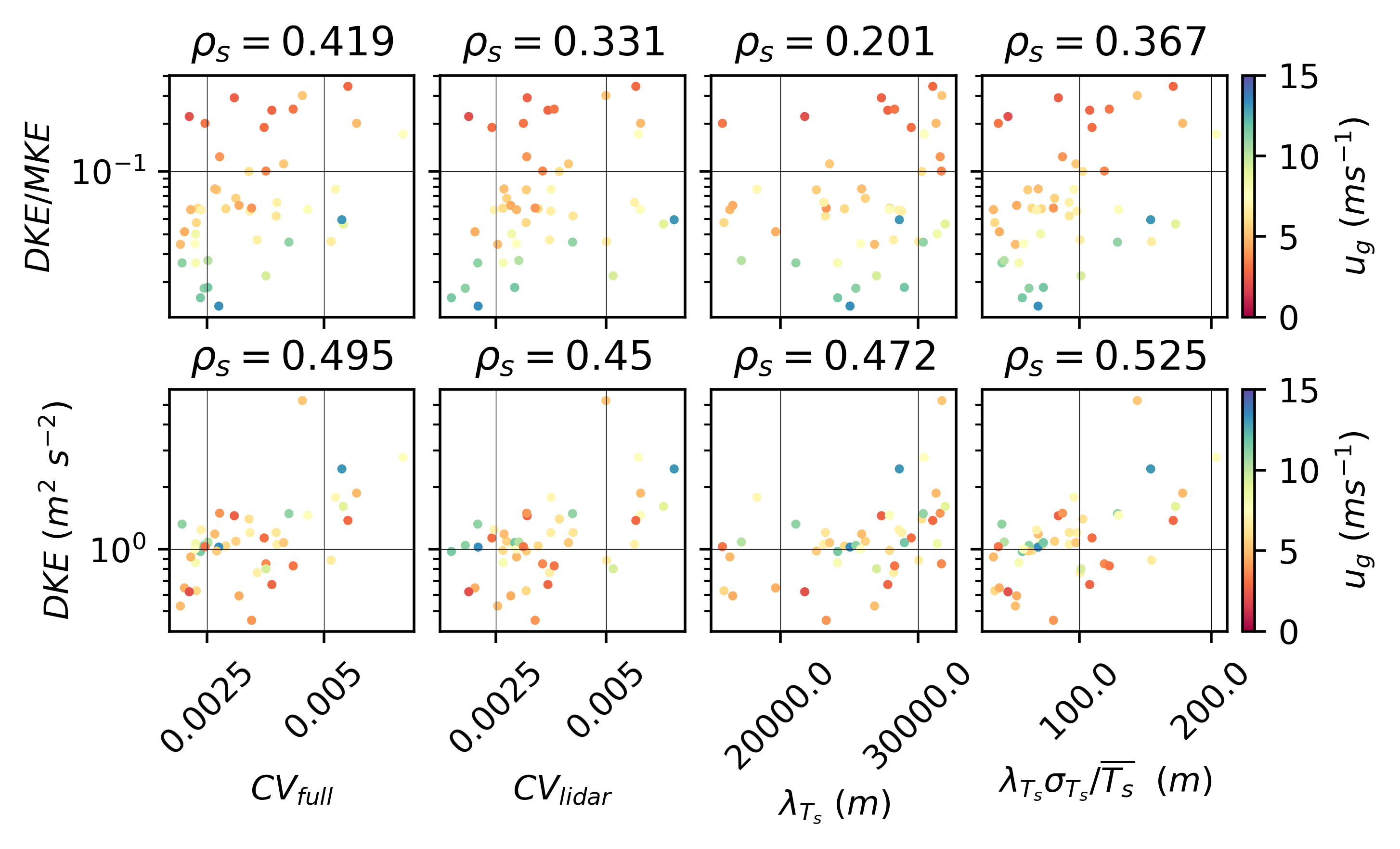}
    \caption{Scatterplots as in figure \ref{fig:filter_1} for the relationship between various heterogeneity parameters and the DKE ratio (top) and DKE (bottom). The four parameters shown, left to right, include the coefficient of variation $CV=\sigma_{T_s}/\overline{T_s}$ for the full GOES LST field ($CV_{full}$), the sample CV from the GOES LST values at the lidar sites only ($CV_{lidar}$), the lengthscale of heterogeneity ($\lambda_{T_s}$), and the primary heterogeneity parameter $\lambda_{T_s}\sigma_{T_s}/\overline{T}_s$. The spearman rank correlation coefficient is shown above each subplot.}
    \label{fig:het_type}
\end{figure}

\begin{figure}
\noindent\includegraphics[width=2in]{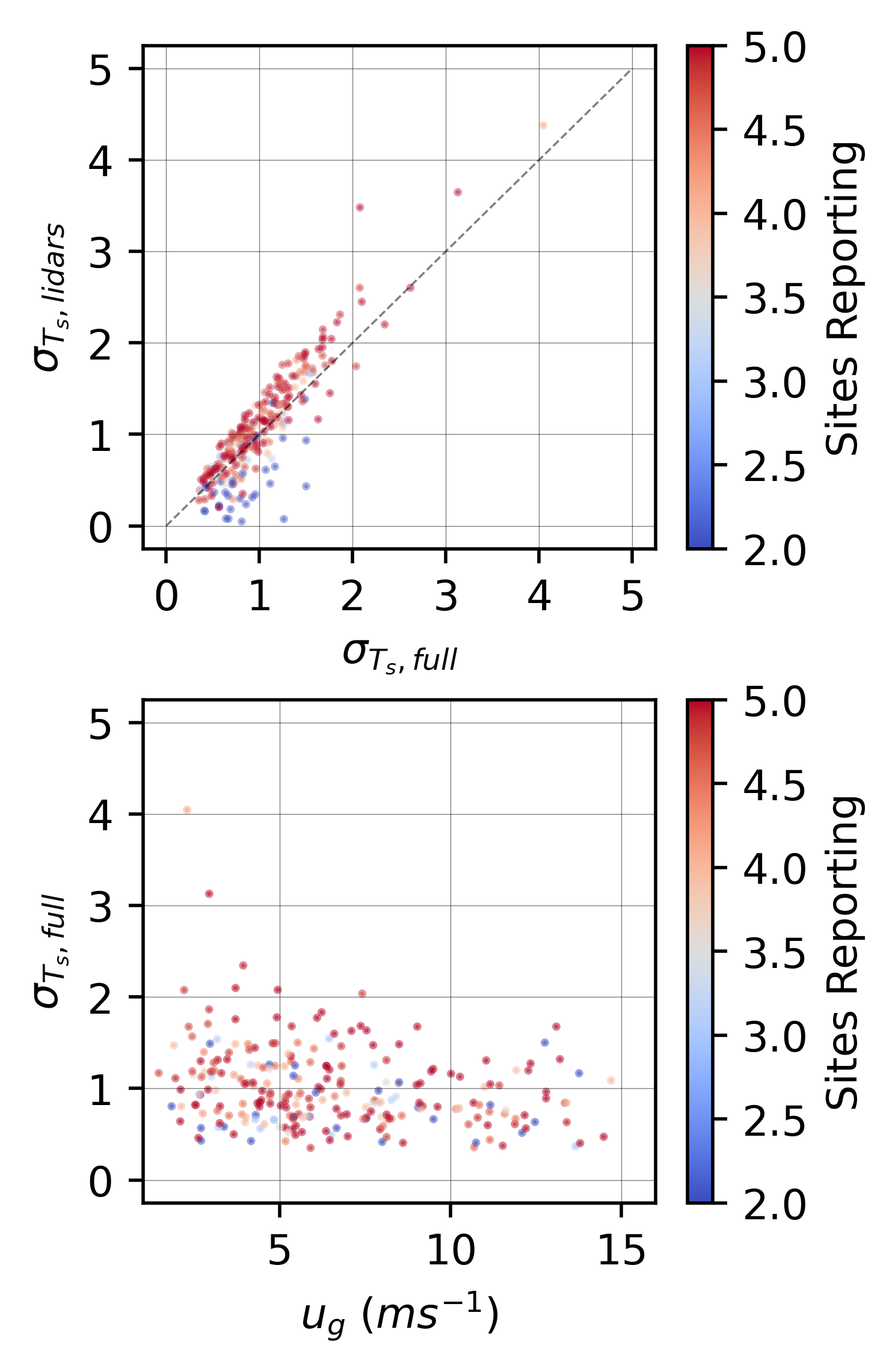}
    \caption{Scatterplots, colored by number of sites reporting of each day, comparing the spatial standard deviation of the full GOES LST field ($\sigma_{T_s,full}$) and the spatial standard deviation from the GOES LST values at the LiDAR sites only ($\sigma_{T_s,lidar}$) (top) and the relationship between the wind speed at $1km$ ($u_g$), and the spatial standard deviation of the full GOES LST field ($\sigma_{T_s,full}$) (bottom).}
    \label{fig:std_v_lidar}
\end{figure}

In addition to sensitivity to atmospheric filters, there will also be some sensitivity to selection of the heterogeneity parameter. In some modeling and observational setups, the lengthscale of heterogeneity $\lambda_{T_s}$ and the coefficient of variation for the full field may be impractical to compute. In figure \ref{fig:het_type}, we explore sensitivity  of land-atmosphere coupling to different metrics to quantify the surface heterogeneity. In general, it appears that multiple different heterogeneity parameters can perform well. Notably, when only the coefficient of variation for the full field is examined ($CV_{full}$), without $\lambda_{T_s}$, performance still appears strong. In the case of the DKE ratio, the correlation is even slightly higher than when using the full heterogeneity parameter. In observational networks, only the temperature values at each LiDAR may be accessible. When only the $CV$ between the 3-5 LiDAR sites is used to quantify heterogeneity performance is weaker than $CV_{full}$ but still strong, showing some promise for using $CV_{lidar}$ in measurement networks when clouds are blocking satellite remote sensing based measurements. Figure \ref{fig:std_v_lidar} directly shows the relationship between the standard deviation of the full LST field and the standard deviation of only the LiDAR sites, providing additional evidence of their close relationship and potential for a small network to generally recover the variability of the temperature field. The lengthscale of heterogeneity has a poor, if any, relationship with the DKE ratio. The correlation is stronger with DKE, however the trend appears approximately flat below 30,000 meters, with the few points above 30,000 meters driving any non-constant relationship absorbed by the correlation coefficient. The importance of this particular metric for lengthscale of heterogeneity, or patch size, is questionable, although it is quite possible that similar metrics on lengthscales orientated along and against the mean wind may show improvements. The expected influence of patch size, or heterogeneity lengthscale, on heterogeneity driven flows is non linear, as previous studies show both minimum necessary values and optimal scales rather than impacts that increase continuously with the lengthscale \cite{lee_effect_2019, avissar_evaluation_1998,patton_influence_2005}. Inclusion of $\lambda_{T_s}$ relative to the boundary layer height may also be a stronger indicator. 

Background velocity has a strong imprint on the DKE ratio via the MKE, as is clear in figures \ref{fig:filter_1} and \ref{fig:het_type}. Additionally, there appears to be a relationship between MKE and the surface heterogeneity that is relatively surprising. Figure \ref{fig:std_v_lidar} shows this more directly; higher spatial standard deviations are more common for low velocities. While much of this manuscript explores land-atmosphere interactions in one direction, the impact land has on the atmosphere, the coupling is assuredly dynamic. Large background velocities are likely to have a homogenizing affect on the surface temperature fields, which may explain the high spatial standard deviations under low velocities and the correlation between 1/MKE and the heterogeneity parameter in figure \ref{fig:filter_1}. The comparison between surface heterogeneity and the DKE ratio, accordingly, will likely have a significant imprint from the MKE's affect on surface patterns, unlike the DKE which shows no clear influence from background velocity. Figures \ref{fig:filter_1}, \ref{fig:filter_2} and \ref{fig:het_type} show a consistent picture that when some steps are taken to isolate for large scale flows and the heterogeneity is oriented perpendicular to the mean wind, DKE shows a strong imprint from the surface heterogeneity and may serve as an observational measure of heterogeneity driven mesoscale atmospheric flows. 

\subsection{Sensitivity to Network Configuration} \label{sec:results_network}
The relatively strong correlations, despite using only 3 to 5 profile measurements for velocity across a 100km by 100km domains, is very promising for future work leveraging LiDAR and DKE to have a first order indication of heterogeneity driven flows. The exact configuration of the tower network will have an influence on the accuracy of these measurements for DKE and DKE ratio. Figure \ref{fig:het_type} and figure \ref{fig:std_v_lidar} already show that a few points may be sufficient to capture first order surface heterogeneity, $\sigma_{T_s}$, in a network, although the effectiveness of $\sigma_{T_s,lidar}$ in capturing $\sigma_{T_s,full}$ will certainly vary depending on the LiDAR placement relative to prevalent heterogeneity patterns in other locations around the globe. For the atmosphere, while sensitivity to network size and measurement location is not possible to explore effectively directly in the LiDAR data, such sensitivity is straightforward to examine in LES. In LES, the correlation between 1/MKE and spatial heterogeneity is not prevalent as the coupling is one way (the surface is imposed and does not respond dynamically to the atmosphere). As such we focus on the DKE ratio for this sensitivity analysis. The network analysis will both provide a rough approximation of the accuracy of results in sections \ref{sec:results_sensitivity} and \ref{sec:results_het} and provide information for experimental design in the establishment of other observational LiDAR networks.

\begin{figure}
\noindent\includegraphics[width=\textwidth]{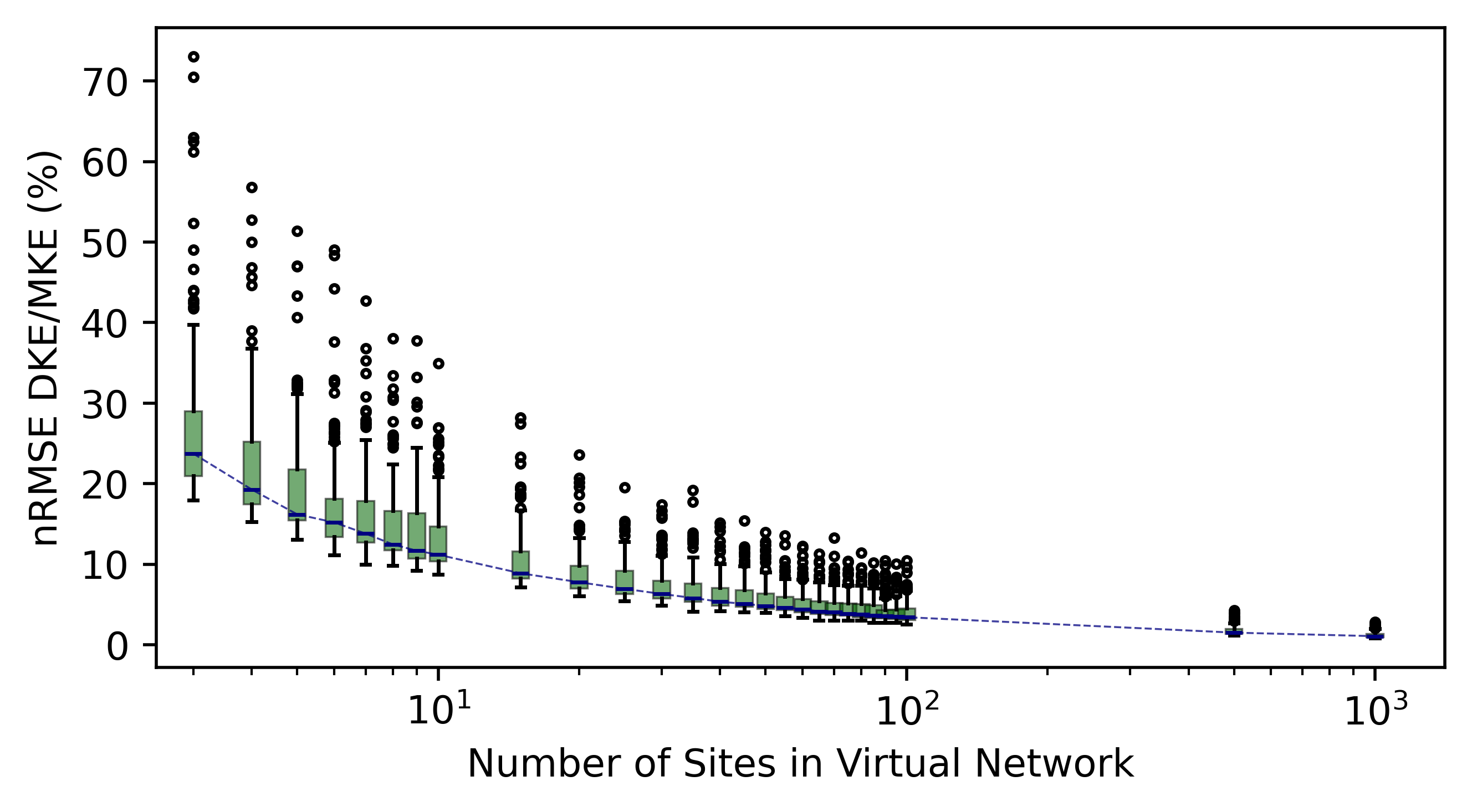}
    \caption{Sizes of virtual LiDAR networks compared to the normalized RMSE of the DKE ratio compared to the DKE ratio from the full LES field. Each boxplot shows the distribution of nRMSE values across 100 random virtual networks of a given size. A dotted line is used to connect the median values of each boxplot.}
    \label{fig:les_rmse}
\end{figure}

\begin{figure}
\noindent\includegraphics[width=\textwidth]{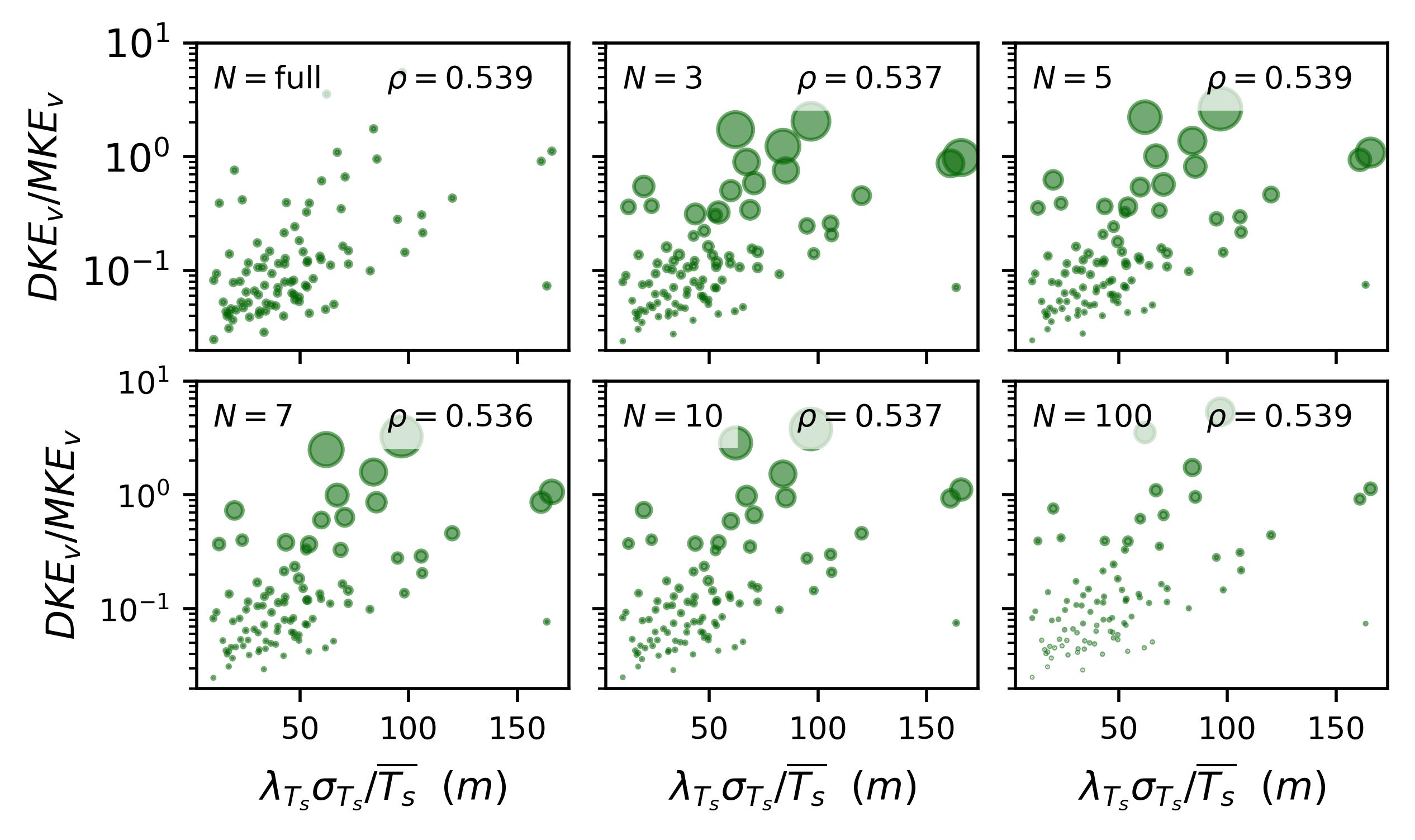}
    \caption{Scatterplots comparing the heterogeneity parameter $\lambda_{T_s}\sigma_{T_s}/\overline{T}_s$ to the DKE ratio across 92 LES days for different sizes of virtual LiDAR networks. Network size is shown in the top right of each subplot, and the spearman rank correlation coefficient shown in the top right. Each point in the scatterplot shows the median DKE ratio across 100 random network configurations. The size of each point in the scatterplot is proportional to the standard deviation of the distribution of DKE ratios across the 100 network configurations for that day.}
    \label{fig:les_scatter}
\end{figure}

In figures \ref{fig:les_rmse} and \ref{fig:les_scatter}, we examine 3,000 virtual networks of LiDAR measurements in the LES across 92 simulation days; a total of 30 different network sizes with 100 geographic configurations for each size. Figure \ref{fig:les_rmse} shows the ability of networks of different sizes to accurately estimate the DKE of the full domain. Error is shown as a root mean squared error normalized by the DKE ratio of the full field (nRMSE). Across the 100 networks, median error is just above 20\% for a network of three virtual LiDAR sites, the minimum size for the filter applied in section \ref{sec:results_sensitivity}. For the largest network size in the real LiDAR observations, 5, median error drops to 15\%. For networks larger than five sites, the marginal improvement in the median error becomes small ($<1\%$ per virtual LiDAR added). While the median is fairly small, there are notable deviations of large errors that continue to be $>20\%$ until the network size is larger than 30 sites. For some site configurations, and surface geometries on certain days, small networks are unable to capture the DKE ratio effectively without large network sizes. Despite this, the relatively low median error provides some confidence for the ability of small network sizes to capture these phenomena statistically, and quantifies the uncertainty that we can expect for the LiDAR-based $DKE$ measurements in section \ref{sec:results_sensitivity} and \ref{sec:results_het}. 

Figure \ref{fig:les_scatter} shows relationship between the heterogeneity parameter and the DKE ratio, as well as a rough indication of the uncertainty in the DKE ratio, for the full LES field and a selection of network sizes across the 92 LES days. The correlation found in LES is significantly higher than in the LiDAR measurements, and the relationship between the two appears relatively consistent. There is no filtering of days in the LES. There is some indirect filtering as the simulations were all done over shallow convection days and with periodic boundary conditions, so deep convection, frontal systems and jets are all not expected to influence the DKE measurements. The correlation is also very consistent across network sizes, indicating that even if there is some error and uncertainty in the DKE measurement for small networks, it does not affect the precision significantly, nor the overall trend of DKE ratio indicating the impact of heterogeneity on the atmosphere. 

\section{Discussion}
The work shows a significant potential for the ability of DKE measured directly from LiDAR to serve as an indicator of heterogeneity driven flows, especially when isolating for weak synoptic forcing and flows perpendicular to the heating gradient. Additional work can certainly refine and expand this analysis into new areas. 

In particular, a more in depth analysis into defining the critical parameters defining the spatial geometry would be valuable; a directional lengthscale of heterogeneity aligned perpendicular to the mean wind may provide a tighter atmospheric connection. The $\alpha$ parameter, which captures some directionality, makes an implicit assumption that the heterogeneity is monotonically increasing in one direction. If the pattern of heterogeneity is repeating, as may be the case if the domain covers a series of ridges and valleys, the direction of the gradient in heating will be unable to capture the pattern and $\alpha$ may be non-representative. Alternative metrics that sum up high-gradient boundaries over the domain, rather than having one directionality for the whole domain, such as those in \citeA{waterman_twocolumn_2024} may be more appropriate. Additionally, the impact of the resolution of the surface heating information, which by its nature filters the heterogeneity of certain scales, and the temporal persistence of the patterns of surface heating cannot be ignored. Figure \ref{fig:std_v_lidar}b also illustrates the importance of considering the two-way land-atmosphere coupling in any analysis.

The analysis also opens the door to new opportunities to explore additional conditions and phenomena, and verify existing findings from LES observationally. Boundary layer height is expected to have a significant affect on how surface heterogeneity impacts atmospheric flows. Incorporation of additional datasets to test if the boundary layer height remains a critical parameter in observations of DKE and heterogeneity driven flows is important and was not explored in this work due to limitations of the available data. The boundary layer height could also serve as an alternative, less arbitrary limit of integration for $DKE_v$ and $MKE_v$ as opposed to the 500 meter integration limit chosen based on the vertical profiles in figure \ref{fig:dke_prof}a,c and figure \ref{fig:ratio_prof}.  While the work focuses on dispersive kinetic energy as a measure of the flow, using dispersive fluxes to explore the vertical transport more direclty may be possible as well. Dispersive heat and moisture fluxes have been examined for heterogeneity driven flows previously in LES \cite{paleri_impact_2025,margairaz_surface_2020-1,akinlabi_dispersive_2022} and with observations of very dense flux tower networks \cite{morrison_heat-flux_2022}. Similar dispersive fluxes could be successfully computed if Doppler LiDAR is collocated with Raman LiDAR which can measure similar vertical profiles of water vapor and temperature. It remains unclear if the small network size would be as sufficient for accurate determination of dispersive fluxes as it appears for DKE/MKE in figure \ref{fig:les_rmse}. At the surface, results in figures \ref{fig:het_type} and figure \ref{fig:std_v_lidar}a imply that using only a small number of surface temperature readings collocated with LiDAR can capture some key heterogeneity driven inputs. The success of $\sigma_{T_s,\ lidars}$ in representing the full field in this location implies that analysis could be expanded to include the days where clear sky GOES-LST readings are unavailable. Inclusion of non-clear sky days increases opportunities to understand how spatio-temporal variance of the surface field impacts the DKE as well as how spatial fields respond in convective conditions.

Even a relatively small number of LiDAR sites is sufficient to capture the relationships between surface heterogeneity and DKE with a reasonable degree of certainty in the DKE measurements from the sensitivity analysis in section \ref{sec:results_network}. This is especially promising given a recent push to establish broader LiDAR based measurement networks of the boundary layer \cite{wulfmeyer_new_2018,hohenegger_fesstval_2023}, where this framework can be applied to explore heterogeneity driven flows over diverse land configurations. There remain some limitations in the approach. The Doppler LiDAR instrumentation at ARM-SGP in particular is unable to capture the TKE. Profiles which include TKE would allow for a more complete assessment of equation \eqref{eq:mke_dke_tke}. TKE is a possible alternative normalizing term for $DKE$ in place of $MKE$, considering the $MKE$ potentially introduces the wind velocity as an important term in both quantification of flow and in the spatial heterogeneity as indicated in figure \ref{fig:filter_1} and figure \ref{fig:std_v_lidar}b. An additional challenge in the DKE metric is separating the component of DKE driven by heterogeneity driven flows from the component attributable to non-local flows such as jets, deep convection, and frontal systems. Future work, parituclarly using co-located LES experiments and LiDAR may be able to address this challenge as well as others in LiDAR-based indicators of heterogeneity driven flows.

\section{Conclusion}
This work examines the potential for examining flows driven by heterogeneity through direct observation as opposed to simulations typically applied to heterogeneous land-atmosphere interactions. The dispersive kinetic energy, DKE, and its ratio with the mesoscale kinetic energy, DKE/MKE, are introduced as metrics to capture the strength and characteristics of the atmospheric flow. The DKE and DKE ratio are computed directly from a LiDAR-based observational network at ARM-SGP, while the characteristics of surface heterogeneity are calculated from satellite based GOES land surface temperature. The DKE and heterogeneity are examined separately and together, and compared qualitatively to findings in LES. Finally, LES experiments are used to illustrate how representative DKE and DKE ratio from networks of multiple LiDAR are for the full flow field.

The results show that DKE and DKE ratio are strong indicators for heterogeneity driven flows. The DKE appears to best capture flows driven by the surface heterogeneity when the influence of large scale flows, such as frontal systems, deep convection, and jets, are accounted for. For the real, measured heterogeneity the exact quantification of the geometry of the surface that matters for atmospheric impact is complex, with a number of different surface parameters appearing to be related to the magnitude of DKE. It is clear that the orientation of the surface heterogeneity relative to the background wind is one of the most critical parameters for predicting whether surface heterogeneity impacts the atmosphere. As such, deeper investigations into alternative metrics to quantify the surface heterogeneity may further improve the results. Network sensitivity analysis suggests that even small LiDAR-based networks can capture DKE with reasonable uncertainty, and that the first-order representation of heterogeneity, spatial standard deviation, can be captured reasonably well without the full surface field and with only surface temperature at each of the LiDAR sites. While this work does not say with certainty that DKE directly measures heterogeneity driven circulations or internal boundary layers, it is clear from this analysis that DKE captures heterogeneity driven flows when they occur. The behavior of DKE in LES, where secondary circulations can be clearly observed, qualitatively matches the behavior of DKE from LiDAR. The LiDAR based DKE, and its interactions with heterogeneity, also reproduce expected behavior based on decades of LES experiments of heterogeneity driven breezes, circulations and boundary layers. Many questions still remain that can be answered by taking advantage of newly established networks and LES experiments. LiDAR-based measurements of dispersive kinetic energy have strong potential as a way to explore changes to boundary layer flow from surface heterogeneity directly with observations. This analysis framework can be leveraged for improved understanding, and weather and climate model representation, of land-atmosphere interactions in heterogeneous systems.

\section*{Open Research Section}
The LiDAR data is available from the Department of Energy through the ARM data discovery portal \cite{shippert_newsom_riihimaki_zhang}. The GOES LST data is publicly available from NOAA's Comprehensive Large Array‐Data Stewardship System (CLASS). The LES model code, inputs \cite{simon_2023_8240267} and summary of outputs \cite{simon_2023_8241941} are also available. The software to create the figures and process the data contained within this work and compute DKE is also available via Zenodo \cite{tyler_waterman_2025_17553087}.

\acknowledgments
The authors thank the various agencies that have supported this work, including the National Science Foundation Division of Atmospheric and Geospace Sciences (NSF-AGS) Postdoctoral Fellowship (Award 2412560; PI Dr. Tyler Waterman) and the Department of Energy through "Observing and understanding the role of surface thermal heterogeneity in mesoscale circulations over AMF3 BNF: Implications for land-atmosphere interactions" (Award DE-SC0025501; PI Dr. Nathaniel Chaney)

%
%

\bibliography{dke}

%
%
%
%
%

\end{document}